\documentclass[pre,twocolumn,aps,showpacs]{revtex4}
\usepackage{dcolumn}
\usepackage{graphicx}
\usepackage{graphicx,amssymb,amsmath}
\usepackage{natbib}
\usepackage{bm}
\def\be{\begin{equation}}
\def\ee{\end{equation}}
\def\ba{\begin{eqnarray}}
\def\ea{\end{eqnarray}}

\begin{document}

\title{Non-equilibrium mechanics and dynamics of motor-activated gels}

\author{F.C.~MacKintosh$^{1,2}$ and A.J.~Levine$^{3}$}

\affiliation{$^{1}$Department of Physics \& Astronomy, Vrije Universiteit, 1081HV Amsterdam, The Netherlands\\
$^{2}$The Kavli Institute for Theoretical Physics, University of California, Santa Barbara, CA 93106\\
$^{3}$Department of Chemistry \& Biochemistry and the California Nanosystems Institute,
University of California, Los Angeles, CA 90095\\
}
\date{\today}
\begin{abstract}
The mechanics of cells is strongly affected by molecular motors that generate forces in the cellular cytoskeleton.  We develop a model for cytoskeletal networks driven out of equilibrium by molecular motors exerting transient contractile stresses. Using this model we show how motor activity can dramatically increase the network's bulk elastic moduli. We also show how motor binding kinetics naturally leads to enhanced low-frequency stress fluctuations that result in non-equilibrium diffusive motion within an elastic network, as seen in recent \emph{in vitro} and \emph{in vivo} experiments. \end{abstract}


\maketitle
The mechanics of living cells are largely governed by the \emph{cytoskeleton},
a complex network of filamentous protein aggregates and various specialized
proteins and enzymes that couple the filaments together and generate
forces\cite{Alberts-1994-26}. As materials, \emph{in vitro} networks of
cytoskeletal filaments have been shown to have unusual mechanical properties,
including a highly non-linear elastic
response\cite{Mackintosh-1997-20,Gardel-2004-2,Storm-2005-5,Bausch-2006-6,Chaudhuri-2007-27}
and negative normal stresses\cite{Janmey-2007-21}. Cytoskeletal networks {\em in vivo},
however, are far from equilibrium materials, due in large part to molecular
motors that exert internal forces within the networks. This presents a
challenge for quantitative statistical/thermodynamic modeling. Recent
studies of \emph{in vitro} networks that include molecular motors have
shown nearly a 100-fold stiffening of the networks due to motor
activity, as well as pronounced low-frequency, non-equilibrium
fluctuations\cite{Mizuno-2007-1}. Here, we develop a model for
such active gels that can explain both the strong stiffening of
networks with motor activity, as well as the large non-equilibrium
fluctuations at low frequencies. We also show how motor (un)binding
kinetics naturally leads to a very simple and general form of stress
fluctuations and diffusive-like motion, which are consistent with
observed non-equilibrium dynamics in living cells\cite{Lau-2003-10,Brangwynne-2007}.
This model can form the basis for quantitative design principles
for creating synthetic polymeric materials with tunable elastic
properties and muscle-like activation.

\begin{figure}[ht]
\centering
\includegraphics[width=6.5cm]{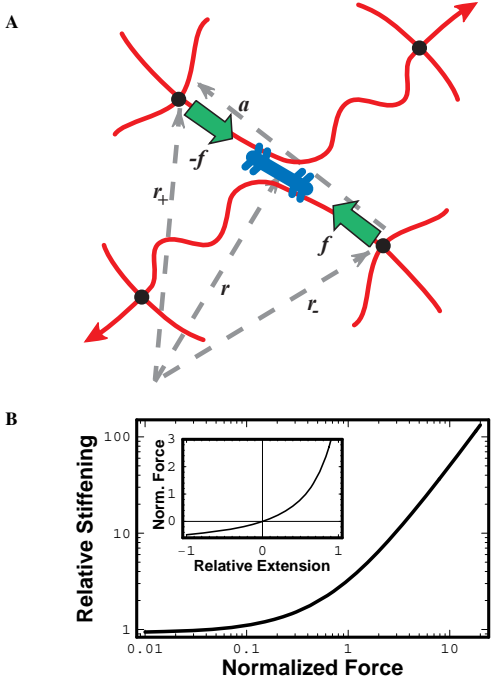}
\caption{(color online) (A) Schematic diagram of contractile motor activity
in a network. A myosin minifilament (blue) slides two network
filaments (red) past each other, generating an equal and opposite
pair of forces (green arrows). (B) Plot of the predicted relative
stiffening of a semi-flexible network as a function of (normalized)
motor-induced tension. The inset shows the nonlinear force-extension
relation of a single semi-flexible
filament\protect{\cite{Mackintosh-1995-4,Gardel-2004-2,Storm-2005-5}}.}
\label{Schematic}
\end{figure}

\emph{Active} solutions consisting of polymers and
motors motors constitute a strikingly new
kind of material that can actively change/adapt its macroscopic
mechanical properties due to small-scale motor
activity that drives relative sliding of polymers past each other\cite{Liverpool-2001-38,Humphrey-2002-9,LeGoff-2002-37,Kruse-2005-39,Liverpool-2006-28}.
In permanently cross-linked networks, however, such motor activity can produce tensile stresses\cite{Mizuno-2007-1}.
This muscle-like contraction is sketched in Fig.\ \ref{Schematic}A. It is well known that
single semi-flexible polymers stiffen under
extension\cite{Bustamante-1994-3}, and that this can result
in macroscopic stiffening of networks under external
strain\cite{Mackintosh-1995-4,Gardel-2004-2,Storm-2005-5}.
This effect can also account for the observed dramatic stiffening
of active networks\cite{Mizuno-2007-1,Koenderink-2007}. Assuming an average
state of tension in the network strands due to motor activity,
we can calculate the expected degree of network stiffening as
follows. The tension $\tau$ in a single filament is calculated
as a function of longitudinal extension $\ell$ as in
Ref.\ \cite{Mackintosh-1995-4}, from which an effective spring
constant $K=d\tau/d\ell$ is calculated. In the nonlinear
regime, this increases as $K\propto\tau^{3/2}$ \cite{Gardel-2004-2}.
The network modulus is given by $G=\frac{1}{15}\rho\ell_c K$,
where $\rho$ is the density (length per volume) of polymer,
and $\ell_c$ is the distance between cross-links\cite{Gittes-1998-8,Morse-1998-32}.
The predicted stiffening is shown in Fig.\ \ref{Schematic}B,
where the filament tension has been normalized by the
characteristic tension $\tau_0=kT\pi^2\ell_p/\ell_c^2$ required to pull out the
fluctuations on a filament of length $\ell_c$ in the network. Here,
$\ell_p$ is the persistence length. For a network of actin filaments,
such as in Mizuno et al.\cite{Mizuno-2007-1}, where $\ell_p=17\mu$m
and $\ell_c\simeq3\mu$m, this characteristic average tension is of
order 0.1pN, meaning that a tension of just a few pN, which is
easily reached by myosin motors, can lead to the observed $100$-fold stiffening
of active networks.

\begin{figure}[ht]
\centering
\includegraphics[width=6.cm]{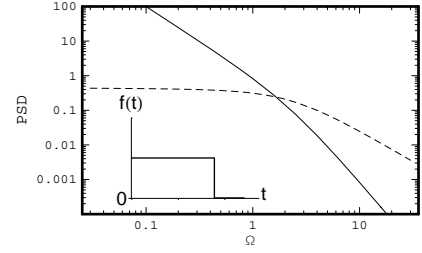}
\caption{The displacement power spectral density (PSD) in an active gel. Here, frequency is measured in terms of $\Omega=\omega\Gamma/B$. The thermal PSD (dashed line) shows a plateau at low frequencies. Thus, the active component of the PSD dominates at low frequencies, while the thermal PSD is expected to dominate at high frequencies.
(Inset) Schematic of the time-dependent force due to molecular motor activity.}
\label{PSDs}
\end{figure}

The quasistatic picture sketched in Fig.\ \ref{Schematic}A
shows a motor (myosin minifilament) generating a pair of equal and
opposite forces   $\mp\vec f$ applied at points
$\vec r_\pm=\vec r\pm\vec a/2$, separated by  $\vec a$. We expect $a$ 
to be a few microns in an \emph{in vitro} network.  Since actin filaments are not able to support
compressive loads over this distance, the resulting force
dipole is contractile: the points are pulled together by a
sort of muscle-like activity. While individual myosin motors are non-processive and
are incapable of persistent, directed motion, they self-assemble into minifilaments,
which are processive. These minifilaments still have a finite duty ratio. When they
unbind the tension is instantaneously released, as sketched in the inset of
Fig.\ \ref{PSDs}\cite{Mizuno-2007-1}. Such a step-like force $f(t)$ corresponds to a power spectrum of force fluctuations that varies as $\omega^{-2}$, proportional to the square Fourier transform of $f$.

As we show, this physical picture of step-like contractile forces naturally leads to
non-equilibrium fluctuations that dominate only at low frequencies, as sketched in Fig.\ \ref{PSDs}.
Surprisingly, this generates motion that appears to be diffusive: $\langle|x(t)-x(0)|^2\rangle\sim Dt$, but
occurring in an \emph{elastic} material. The effective diffusion constant $D$
is controlled by motor activity and not temperature.
Using well-established viscoelastic
properties of cross-linked F-actin networks\cite{Gittes-1998-8,Morse-1998-32},
we find distinct regimes of both thermal and athermal
(motor-induced) fluctuations sketched in Fig.\ \ref{PSDs},
which are consistent with the observations both \emph{in vivo}
\cite{Lau-2003-10} and \emph{in vitro}\cite{Mizuno-2007-1}.

To model the active gel we use a continuum description for a viscoelastic homogeneous and isotropic medium,
but in which the motor activity couples to this medium as illustrated in Fig.\ \ref{Schematic}A.
\begin{figure}[b]
\centering
\includegraphics[width=7.cm]{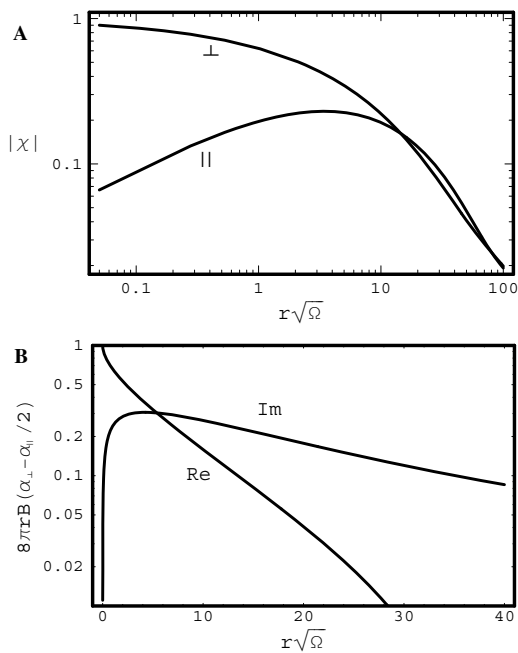}
\caption{(A) Graphs of the spatial dependence of the
longitudinal parts of the parallel ($\parallel$)
and perpendicular ($\perp$) response functions (Eqs.\ (2) and (3)).
The effect of compression of the network on the response
functions can be reduced to a universal form when plotted against
the dimensionless quantity $r\sqrt{\Omega}=r\sqrt{\omega\Gamma/B}$,
demonstrating the diffusive nature of the propagation of the network
density mode. (B) The effect of network compression can be isolated
in experimental data by examining the difference in the parallel and
perpendicular response functions given in Eq.\ (4). Here we plot the
predicted form of the real (Re) and imaginary (Im) parts of that
difference vs. the dimensionless variable $r\sqrt{\Omega}$.
}\label{TheChis}
\end{figure}
For \emph{in vitro} networks such as in Ref.\ \cite{Mizuno-2007-1}, the distance between cross-links, and thus $a$ is
expected to be of order 3-10$\mu$m. On this scale, we can model the action of a motor as the introduction of a pair for equal and opposite applied forces in the (visco-)elastic continuum. The resulting displacement field $u_i$ at position $\vec r_0$ of the network we describe by a linear response function $\alpha_{ij}$ depending on position and frequency as
\be
u_i\left(\vec r_0,\omega\right)=\left[\alpha_{ij}\left(\vec r_0-\vec
r_-,\omega\right)-\alpha_{ij}\left(\vec r_0-\vec r_+,\omega\right)\right]f_j(\omega),
\ee
using the fact that the motor-generated forces $\mp \vec{f}$ are equal and opposite.
Stability also requires that $\vec{f}$ and $\vec{a}$ be parallel. The response function to a point force $\alpha_{ij}$ can be written in terms of $\alpha_\parallel$ and $\alpha_\perp$, where $\alpha_{ij}\left(\vec r\right)=\hat r_i\hat
r_j\alpha_\parallel\left(r\right)+\left(\delta_{ij}-\hat r_i\hat
r_j\right)\alpha_\perp\left(r\right)$.

\begin{figure}[b]
\centering
\includegraphics[width=6.5cm]{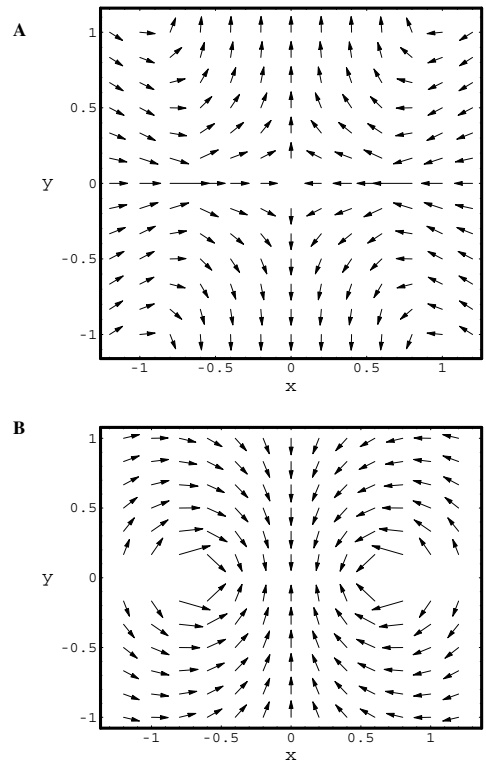}
\caption{(A) The displacement vector field of an incompressible
network shown in a plane passing through the two force centers
for a contractile motor acting at the origin. The
forces are applied symmetrically at points $(\pm3/4,0)$ and are
each directed towards the origin. (B) The network displacement
field for the compression mode shown in the limit of
low frequency or weak hydrodynamic coupling ($\Gamma\rightarrow0$).
Again the forces are applied symmetrically at points $(\pm3/4,0)$
and are each directed towards the origin. The resulting displacement
field induces network density variations in the material.}
\label{VecField}
\end{figure}
We calculate these two response components within a
two-fluid approximation, in which the cytoskeletal filaments are
treated as a porous elastic network immersed in a viscous
solvent\cite{Brochard-1977-35,Milner-1993-34,Gittes-1997-19,Levine-2000-16}.
Here, the network displacement $u$ and solvent velocity $v$ satisfy the coupled equations
\be
0=\mu\nabla^2\vec u+(\mu+\lambda)\vec\nabla(\vec\nabla\cdot\vec u)+
\Gamma\left(\vec v-\frac{d \vec{u}}{dt}\right)+\vec f_n,
\ee
%
%
\be
0=\eta\nabla^2\vec v-\vec\nabla P-\Gamma\left(\vec v-\frac{d\vec{u}}{dt}\right)+\vec f_s,
\ee
where $\mu$ and $\lambda$ are Lame coefficients, $\eta$ is the solvent viscosity,
and the forces $f_{n,s}$ represent the forces on the network and solvent,
respectively. Given a meshwork with a pore size $\xi$, the coupling $\Gamma$
is expected to be of order $\eta/\xi^2$. These are solved for the
response of the combined system to an applied point force. The resulting
response functions are given by
\be
\alpha_\parallel\left(r,\omega\right)= \frac{1}{4\pi
rG\left(\omega\right)}\left[1+\frac{G\left(\omega\right)}{B\left(\omega\right)}
\chi_\parallel\left(r\sqrt{\Omega}\right)\right],
\ee
and
\be
\alpha_\perp\left(r,\omega\right)= \frac{1}{8\pi
rG\left(\omega\right)}\left[1+
\frac{G\left(\omega\right)}{B\left(\omega\right)}
\chi_\perp\left(r\sqrt{\Omega}\right)\right],
\ee
where
$\chi_\perp\left(x\right)={2i}
\left[1-\left(1+x\right)e^{-x\sqrt{-i}}\right]/x^2$ and
$\chi_\parallel\left(x\right)=e^{-x\sqrt{-i}}-\chi_\perp\left(x\right)$.
Here, $G$ is the shear modulus and $B=\frac{2(1-\sigma)}{1-2\sigma}G$
is the longitudinal modulus, where $\sigma$ is the Poisson ratio,
and $\Omega=\omega\Gamma/B$. This coupling can
be understood in terms of the solvent flow through the highly porous
gel: rapid solvent flow through the filament mesh gives rise to
large shear stresses, effectively dragging the network with the solvent.
This drag prevents the large-scale relative motion of the network and solvent
beyond a range of order $\Omega^{-1/2}$.
On larger length scales $r$ or at higher frequencies $\omega$, the
drag effectively inhibits the relative motion of solvent
and network so that for $r\sqrt{\Omega}\gg 1$, the combined network
and solvent act as a single incompressible material\cite{Gittes-1997-19,Levine-2000-16}, and
$\chi_{\parallel,\perp}$ both vanish (Fig.\ \ref{TheChis}A).
Here, the response of the medium is purely \emph{transverse}
(the displacement vector field is divergenceless) and is given
by the generalized Oseen tensor, given by
leading terms in square brackets above\cite{Levine-2000-16}. The corresponding volume-preserving
flow response of an incompressible gel when subject to a symmetric pair
of point forces is shown in Fig.\ \ref{VecField}A.

In this incompressible case, the displacement field
$u(\omega)$ of the network resulting from motor activity varies with
an overall frequency dependence proportional to the ratio of
the force $f(\omega)$ to the shear modulus $G(\omega)$, according
to Eqs.\ (1-3). Thus, we find for the model illustrated in
Fig.\ \ref{Schematic}A that
$\langle|u(\omega)|^2\rangle\propto\langle|f(\omega)|^2\rangle/|G(\omega)|^2\propto|\omega G|^{-2}$.
Cross-linked biopolymer networks typically exhibit a constant or
weakly frequency-dependent elastic regime as a function of frequency.
Here, we expect to see $\langle|u(\omega)|^2\rangle\propto\omega^{-2}$,
which is consistent with recent displacement fluctuations observed in
cells\cite{Lau-2003-10}, and which corresponds to diffusive motion.
At higher frequencies, such networks typically
exhibit a power-law increase in the shear modulus with
frequency\cite{Gittes-1997-19,Gittes-1998-8,Morse-1998-32}, in
which $G\propto\omega^{3/4}$. In this frequency regime stain fluctuations in
the active gel take the form $\langle|u(\omega)|^2\rangle\propto\omega^{-7/2}$, as
shown in Fig.\ \ref{PSDs}. For comparison, the equilibrium
thermal fluctuations for such a network are shown as the dashed line.
At low frequencies the motor-driven fluctuations will dominate
over the ever-present thermal fluctuations, consistent with the results
of both Lau et al.\cite{Lau-2003-10} and Mizuno \emph{et al}.\cite{Mizuno-2007-1}.

Since biopolymer and cytoskeletal
networks are generically porous with pore
sizes of order $1\mu$m, they can deform \emph{compressibly}.
This density mode, however, is strongly suppressed by drag at
high enough frequencies. The loss of the density mode at high frequencies
is illustrated in Fig.\ \ref{TheChis}A, where the effects of finite
compressibility, represented by  $\chi_{\parallel,\perp}$,
vanish at high frequency. Although the basic physics of these
effects have been discussed before for both flexible polymer
systems\cite{Brochard-1977-35,Milner-1993-34} and semi-flexible
biopolymer systems\cite{Gittes-1997-19,Levine-2000-16}, there
has been no direct experimental observation of these compressibility
effects in porous biopolymer systems.

We can isolate the effects of the network compressibility by the examining the combination
\be
\alpha_\perp(r,\omega)-\frac{1}{2}\alpha_\parallel(r,\omega)=
\frac{\left[\chi_\perp\left(r\sqrt{\Omega}\right)
-\frac{1}{2}\chi_\parallel\left(r\sqrt{\Omega}\right)\right]}{8\pi
rB\left(\omega\right)},
\ee
which is plotted in Fig.\ \ref{TheChis}B. This measurable
combination of response functions strictly vanishes in the
incompressible limit. This, along with the specific combined
$r$ and $\omega$ dependence, may permit the first direct measurement
of compressibility effects that are expected to be characteristic
of biopolymer/cytoskeletal networks. Furthermore, the flow/displacement
field corresponding to this compressible mode (shown in Fig.\ \ref{VecField}B
in the limit $\Gamma\rightarrow0$)
strongly differ from the case of an incompressible system (Fig.\ \ref{VecField}A).
Here, the \emph{longitudinal} (irrotational displacement field)
contributions to the response function are $\alpha_\parallel^{(L)}=0$
and $\alpha_\perp^{(L)}=1/(8\pi rB)$. The difference in
spatial structure of these strain fields may also be used to experimentally
identify the effects of compression.

To consider the effect of multiple contractile events within the medium, we can
represent the resulting displacement field at the origin $u_i$ at
by a sum
\be
u_i=\sum\Delta\alpha_{ij}\left(\vec r,\vec a\right)\hat a_jf,
\ee
where $\Delta\alpha_{ij}\left(\vec r,\vec a\right)=
\alpha_{ij}\left(\vec r-\vec a/2\right)-\alpha_{ij}\left(\vec
r+\vec a/2\right)$ is the response to a contractile force pair.
We suppress the frequency dependence.
This sum represents the combined effect of
temporally uncorrelated contractile events occurring
homogeneously throughout the medium.
This assumption remains valid provided that the events
rarely occur with a separation of order $a\sim \ell_c$
during the typical processivity time $t_0$. Such a
sum or average has been performed in calculating the PSD in
Fig.\ \ref{PSDs} for the case of an incompressible network.
In this case the scaling described
above is a good approximation.

This model shows how motor activity within a semi-flexible gel,
together with the well-established non-linear response of such
networks leads to a strong stiffening of the network,
and that this stiffening increases more than linearly with the motor
force. This can account for the recently observed nearly 100-fold
network stiffening with motor forces of order 1-10 pN\cite{Mizuno-2007-1}.
Furthermore, the (un)binding kinetics of the motors naturally
leads to a specific characteristic time dependence of the
force fluctuations in active gels. Given a finite processivity time $t_0$ over which minifilaments remain bound and generate force, the unbinding results in $1/\omega^2$ force fluctuations for frequencies $\omega>1/t_0$. This spectrum is a direct result of the expected sharp time dependence of motor unbinding, and is insensitive to slow variations of force during motor motion. For frequencies $\omega>1/t_0$, the divergence of the force spectrum will be suppressed.
Our model, is for uncorrelated
motor activity, in that the total fluctuations can be represented
as a sum of independent fluctuations due to individual motor
force generation and unbinding. At sufficiently high motor densities,
one might expect cooperativity of motor activity, whose
consequences can be studied in extension of the present model.

This work was supported in part by the (Netherlands) Foundation
for Fundamental Research on Matter (FOM), NSF Materials
World Networks (grant no.\ DMR-0354113), and the NSF
through the Kavli Institute for Theoretical Physics.
The authors thank J.\ Crocker, A.\ Grosberg, A.\ Lau,
T.\ Lubensky, D.\ Mizuno, M.\ Rubinstein,
and C.\ Schmidt for helpful discussions.


\begin{thebibliography}{99}

\bibitem{Alberts-1994-26}B.\ Alberts, et al., {\em Molecular Biology of the Cell, 3rd edition}
(Garland, New York, 1994). 

\bibitem{Mackintosh-1997-20}F.C.\ MacKintosh and P.A.\ Janmey, Current
Opinion in Solid State \& Materials Science \textbf{2}: 350 (1997).

\bibitem{Gardel-2004-2}M.L.\ Gardel, et al., Science \textbf{304}: 1301 (2004).

\bibitem{Storm-2005-5}C.\ Storm, et al., Nature \textbf{435}: 191 (2005).

\bibitem{Bausch-2006-6}A.R.\ Bausch and K.\ Kroy, Nature Physics \textbf{2}: 231 (2006).

\bibitem{Chaudhuri-2007-27}O.\ Chaudhuri, S.H.\ Parekh and D.A.\ Fletcher, Nature \textbf{445}: 295 (2007).

\bibitem{Janmey-2007-21}P.A.\ Janmey, et al., Nature Materials \textbf{6}: 48 (2007).

\bibitem{Mizuno-2007-1}D.\ Mizuno, C.\ Tardin, C.F.\ Schmidt and F.C.\ MacKintosh, Science \textbf{315}: 370 (2007).

\bibitem{Lau-2003-10}A.W.C.\ Lau, et al., Phys.\ Rev.\ Lett.\ \textbf{91}: 198101 (2003).

\bibitem{Brangwynne-2007}C.\ Brangwynne, F.C.\ MacKintosh, and D.A.\ Weitz, PNAS, in press; C.\ Brangwynne, et al., arXiv:0709.2952.

\bibitem{Liverpool-2001-38}T.B.\ Liverpool, A.C.\ Maggs and A.\ Ajdari, Phys.\ Rev.\ Lett.\ \textbf{86}: 4171 (2001).

\bibitem{LeGoff-2002-37}L.\ Le Goff, F.\ Amblard and E.M.\ Furst, Phys.\ Rev.\ Lett.\ \textbf{88}: 018101 (2001).

\bibitem{Humphrey-2002-9}D.\ Humphrey, et al., Nature \textbf{416}: 413 (2002).

\bibitem{Liverpool-2006-28}T.B.\ Liverpool, Phil.\ Trans.\ A \textbf{364}: 3335 (2006).

\bibitem{Kruse-2005-39}K.\ Kruse, et al., Eur.\ Phys.\ J.\ E \textbf{16}: 5 (2005).

\bibitem{Bustamante-1994-3}C.\ Bustamante, et al., Science \textbf{265}: 1599 (1994).

\bibitem{Mackintosh-1995-4}F.C.\ MacKintosh, J.\ K\"as and P.A.\ Janmey, Phys.\ Rev.\ Lett.\ \textbf{75}: 4425 (1995).

\bibitem{Koenderink-2007}G.H.\ Koenderink, et al., unpublished.

\bibitem{Gittes-1998-8}F.\ Gittes and F.C.\ MacKintosh, Phys.\ Rev.\ E \textbf{58}: R1241 (1998).

\bibitem{Morse-1998-32}D.C.\ Morse, Macromolecules \textbf{31}: 7044 (1998).

\bibitem{Brochard-1977-35}F.\ Brochard and P.G.\ Degennes, Macromolecules \textbf{10}: 1157 (1977).

\bibitem{Milner-1993-34}S.T.\ Milner, Phys.\ Rev.\ E \textbf{48}: 3674 (1993).

\bibitem{Gittes-1997-19}F.\ Gittes, et al., Phys.\ Rev.\ Lett.\ \textbf{79}: 3286 (1997); B.\ Schnurr, F.\ Gittes, F.C.\ MacKintosh and C.F.\ Schmidt, Macromolecules \textbf{30}: 7781 (1997).

\bibitem{Levine-2000-16}A.J.\ Levine and T.C.\ Lubensky, Phys.\ Rev.\ Lett.\ \textbf{85}: 1774 (2000).

\end{thebibliography}
\end{document}